# Casting Polymer Nets to Optimize Noisy Molecular Codes

TSVI TLUSTY

*Department of Physics of Complex Systems, Weizmann Institute of Science, Rehovot, Israel 76100.*

**Abstract**

Life relies on the efficient performance of molecular codes, which relate symbols and meanings via error-prone molecular recognition. We describe how optimizing a code to withstand the impact of molecular recognition noise may be approximated by the statistics of a two-dimensional network made of polymers. The noisy code is defined by partitioning the space of symbols into regions according to their meanings. The "polymers" are the boundaries between these regions and their statistics defines the cost and the quality of the noisy code. When the parameters that control the cost-quality balance are varied, the polymer network undergoes a first-order transition, where the number of encoded meanings rises discontinuously. Effects of population dynamics on the evolution of the code are discussed.

**Introduction**

In the living cell, information is carried by molecules. The outside environment and the biochemical circuitry of the cell churn out fluxes of molecular information that are read, processed and stored in memory by other molecules. The cell's information-processing networks often need to translate a symbol written in one species of molecules into another symbol written in a different molecular language. This requires a code-table that translates between the two molecular languages. Perhaps the best-known example is the genetic code-table that translates sixty-four DNA base-triplets into twenty amino-acids (*1, 2*). One may think of such a code-table as a mapping – a probabilistic one due to the inherent noise – between the space of molecular "symbols", e.g. the base-triplets, and the space of molecular "meanings", e.g. amino-acids. The notion of mapping between two molecular spaces occurs also in biological codes of a much larger scale, for example the transcriptional regulatory network that controls gene expression by DNA-binding regulatory proteins. This network may be seen as a mapping from the space of regulatory proteins to the space of their respective DNA binding sites. Evolution poses the organism with a semantic challenge: its code-tables must assign meanings to symbols in a manner that minimizes the impact of the molecular recognition errors while keeping down the cost of resources that the code-table necessitates. The present work describes a treatment of this biological optimization problem in terms of the statistical mechanics of polymer networks.

In the biophysical reality of the cell, actual polymer networks are essential for the structural stability and motility (*3, 4*). However in the present context of coding, polymer networks are mathematical entities that prove useful for describing the code-table and studying its optimization: Molecular recognition is inherently noisy since it involves energy scales that are not much larger than the typical thermal energy $k_BT$. To reflect these recognition errors, the space of symbols is depicted as a graph in which symbols are vertices and edges connect vertices that are likely to be confused by misreading (Fig. 1). A code-table is then constructed by assigning meanings to each vertex. This can be pictured as coloring the vertices according to their meaning (*2*), which partitions the graph into "islands of meanings". The boundaries between these islands form a network, which can be likened to a self-assembling network of polymers or self-avoiding random walks (Fig. 1). Polymer networks are natural in this context because they are related to the notion of space partitioning that is central to coding theory (*5*). But the resemblance to a polymer network is not merely structural. Optimizing the fitness of the mapping is shown to be equivalent to minimizing the free energy of the polymer network. Such an optimal mapping must balance the three conflicting needs for maximal error tolerance, for maximal diversity and for minimal cost. During evolution, the code-table adapts by altering the network in response to changes in the equipoise of these three evolutionary forces

In the present work we first discuss how the partition of the symbol space by the polymer networks determines the fitness of the code-table, where mathematical definitions of fitness and its determinants, error-load, diversity and cost are given. Next, we discuss one purpose of the present work, which is to show that the fitness function of the code-table corresponds to the free energy of the polymer network. Thus, it suggests that the problem of optimizing the code-table is equivalent to calculating the equilibrium statistics of the polymer network. Then a second purpose is to use this equivalence to examine the code optimization problem in parameter regimes that are hard to access otherwise. In particular, it is used to identify a first-order transition, in which the number of meaning islands changes abruptly in response to varying the error-tolerance/diversity/cost balance. Finally, to put the model within a context of population dynamics we discuss possible aspects of metastability, mutations and genetic drift that may affect the evolution of the code.

## Model and Results

**The fitness of molecular codes.** Information in the cell is recorded in molecules and then retrieved and translated into other molecules by a code-table. To discuss how the organism optimizes such a code-table, we represent the table as an information channel or a mapping that relates a meaning space, in which $n_m$ meanings reside, to a symbol space, which contains $n_s$ symbols (Fig. 1). The code table maps (or encodes) meanings to symbols and thus partitions the space of symbols into meaning islands whose boundaries form a polymer network. In this section, we first calculate the fitness of a code-table as a function of the mapping as specified by the polymer network. The code fitness is composed of contributions due to the diversity of code table, its average error-load and the cost of constructing the molecular coding apparatus. There are two sources of noise in our simple description of the coding system, noise while reading and noise in the mapping. Recognition errors while reading a symbol are represented by the symbol graph whose edges connect symbols that misreading may confuse. The impact of this noise is included in the error-load component of the fitness. Noise in the mapping, which is implemented in error-prone molecular recognizers, is represented as a statistical average over an ensemble of polymer networks. This consequence of recognition noise is included in the cost (which together with the error-load is defined below).

A code-table does not necessarily map all the $n_m$ available meanings to symbols but may possibly map only $n_f \leq n_m$ out of them. The larger is the number of encoded meanings $n_f$ the more diverse is the code. *Diversity* contributes to the fitness of the code, since it increases the chance that when the organism needs to read, write or process a certain meaning it can accurately encode it as one of the available symbols. However, diversity of meanings also increases the probability for misreading errors: To reconstruct the meaning encoded by a certain symbol the organism has to read it. Since the molecular reading apparatus is not perfect it may sometimes confuse this symbol with one of its neighbors in the graph (Fig. 1). If many meanings are encoded then, on average, the meaning islands defined by the network are smaller. In this finer network, the chance of confusing symbols of different meanings is larger, which costs the organism in a higher error-load.

To find the *error-load*, one needs to specify a partition and examine the average chance to cross by misreading the boundaries between meaning islands. We specify a partition by assigning to each edge *i-j* a binary variable $E_{ij}$ that indicates whether the edge is on the boundary between two meaning islands ($E_{ij} = 1$) or inside an island ($E_{ij} = 0$). Misreading along the edge, which confuses *i* with *j*, occurs at a probability $r_{ij}$, while $r_{ii}$ is the probability to correctly read *i*. There may be two possible outcomes of misreading, which depend on the partition into meaning islands: If both symbols are in the same island and are therefore synonymous, then misreading bears no load since the translated meaning does not change. If the symbols reside in two different islands and are therefore non-synonymous, then the fitness of the organism decreases by one fitness unit. The contribution of an *i-j* misreading to the error-load can be therefore expressed as $r_{ij}E_{ij}$. The total error-load is a sum over all edges, $\sum_{i-j} r_{ij}E_{ij}$, which is equal to the chance that, due to misreading, a symbol is translated incorrectly into a wrong meaning.

An evolutionary force that counteracts the need to minimize error-load is the need for diversity. Our model assumes for simplicity that the contribution of diversity to the fitness is linear in $n_f$, the number of encoded meanings. The *quality* $H_E$, of a given network pattern *E*, is then a linear combination of the error-load and the diversity,

$$H_E = \sum_{i-j} r_{ij} E_{ij} - w_D n_f, \tag{1}$$

where the parameter $w_D$ measures the significance of diversity relative to error-load. For the clarity of the physical analogy we use a sign convention in which a code of high quality corresponds to low values of $H_E$. Thus, $H_E$ behaves like an energy which is minimal when the code is optimal. The quality depends on the coloring pattern of the code-table, which determines its error-load and diversity. As illustrated in Fig. 1, each coloring pattern is determined by the network of boundaries between the islands, which is equivalent to a network of living polymers. The quality is governed by the interplay between error-load and diversity: If the reader was ideal $r_{ij} = \delta_{ij}$, then it would have been advantageous to decode as many meanings as there are available symbols, $n_f = n_s$. However, since the molecular reader is not perfect it is preferable to decode fewer meanings to minimize the effect of misreading errors. The quality $H_E$ corresponds to a "microstate" specified by a deterministic network configuration $E$. The probabilistic mapping of the molecular code is a "macrostate", which is represented by an ensemble of such configurations and is calculated below. The code quality $H_C$ is the ensemble average of quality of a over all microstates, $H_C = \langle H_E \rangle$. The need for this averaging is due to the stochastic nature of the molecular recognition interactions that define the mapping from symbols to meanings.

Besides the quality of the code, which combines its error-load and diversity, the code fitness must also account for the *cost* of the molecular machinery that performs the mapping. Molecular codes are physically implemented by recognition interactions between the meanings, the symbols and sometimes other intermediary molecules, such as the tRNA in the genetic code (1). High specificity of recognition improves the quality of the code since it enables more accurate mapping. However, highly specific binding requires a higher binding energy, which in general necessitates larger binding sites. It is plausible to assume that the cost of the code is proportional to the average size of the binding sites and therefore to the average binding energy (6). This is because the cost of synthesizing the molecules and maintaining their genes is proportional their size. To estimate the cost one notes that the binding probability scales like the Boltzmann exponent of the binding energy (in units of $k_B T$), $P_b \sim \exp(E_b)$. It follows that the cost, which is equal to the average binding energy $\langle E_b \rangle$. can be approximated by the average $\langle \ln P_b \rangle$, which is minus the entropy of the mapping meanings to symbols, $-S_C$ (see Methods). This "macrostate" entropy takes into account the ensemble of all possible mappings, which is determined by all possible networks and the number of possible ways to color every such network.

Finally, the overall fitness of the code $F_C$ is estimated by a weighed sum of the quality and the cost, $F_C = H_C - w_C \cdot S_C$, where the parameter $w_C$ measures the significance of the cost with respect to quality. $F_C$ is the free energy of all possible colorings of the code-table and may be derived from the partition function $Z_C$,

$$Z_C = \exp(-F_C / w_C) = \sum_{mappings} \exp(-H_E / w_C). \quad (2)$$

Within this analogy, the quality $H_E$ plays the role of a Hamiltonian and the cost parameter $w_C$ is an effective temperature. The ensemble average in $Z_C$ is due to the probabilistic nature of the molecular mapping, which is measured by the cost weight $w_C$. At high $w_C$, codes are fuzzy and smeared over many network configurations, while at low $w_C$ they are sharper as the mapping is almost deterministic. In principle, one can derive the code-table by performing the summation in Eq. 2. But in practice this is a burdensome task that can be performed only numerically and even this only for codes of limited size. Tractable analytic results exist mostly at the limit of high $w_C$. In this regime, it is known that the code table undergoes a second-order phase transition from a non-coding state of no correlation between meanings and symbols to a coding state, in which such correlation has just emerged (2, 6-8).

The cost-quality balance of the code-table is analogous to the balance in an engineered noisy information channel between the average distortion in the channel $D$, which measures its quality, and the channel's rate $I$, which measures the cost of the channel by estimating how many bits are required to encode one meaning. Rate-distortion theory (9) focuses on the fundamental problem of optimizing a noisy information channel, which can be formalized as the question: what is the minimal rate $I$ required to assure that the distortion in the channel will be less than a certain desired value $D$? This optimal rate-distortion curve is calculated by minimizing a functional $F = D + w_C I$, where $D, I$ and $F$ are analogues of $H_C$ and $-S_C$ and $F_C$, respectively. The "temperature" $w_C = -\partial D/\partial I$, the slope of the optimal curve, measures the increase in quality due to an additional bit of information. In the biological context, $w_C$ is expected to decrease with the complexity the organism and its environment: A complex organism transmits more information. It is therefore in the interest of this organism to pay a larger cost to improve the quality of its codes, or in other words $w_C = \partial H_C/\partial S_C$ is lower. Similarly, a richer environment is also "colder". At low $w_C$ temperature, the quality $H_C$ dominates the free energy $F_C$ and the code-table tends to one of the many minima of $H_E$. Derivation of the optimal code in this regime is difficult, even numerically, due to rugged landscape of $H_E$. As we discuss below, the polymer network formalism offers insight in this regime and, specifically, suggests a first-order coding transition.

**Statistics of polymer networks on the dual symbol-graph.** To formalize the analogy of molecular codes to polymer networks we need to examine the dual of the symbol-graph on which the network resides (Fig. 1). To find the dual one first embeds the symbol graph in a surface (10). In this example, the symbol graph is a hexagonal lattice that is embedded in a torus and the dual is a triangular lattice (for details on finding the dual graph see Methods). It is evident that the vertex-coloring pattern of the symbol graph corresponds to a connected "polymer" network whose monomers are edges of the dual graph (Fig. 1).

By counting the number of edges and vertices in the polymer network one can derive the number of meaning islands $n_f$ in the expression for the quality $H_E$ (Eq. 1). For this purpose, we introduce another binary variable $V_i$ that indicates whether a vertex $i$ of the dual graph is part of the polymer network ($V_i = 1$) or not ($V_i = 0$). Then, the numbers of edges $n_e$, vertices $n_v$, and islands $n_f$, are related through the definition of Euler's characteristic $\chi = n_v - n_e + n_f$ (10),

$$(3) \qquad n_f = \chi + n_e - n_v = \chi + \sum_{i-j} E_{ij} - \sum_i V_i ,$$

The value of $\chi$ is determined by the topology of the surface in which the symbol graph is embedded, for example $\chi = 0$ for the torus in Fig. 1. By substitution of $n_f$ from Eq. 3 into Eq. 1, we find that the code quality is $H_{EV} = \sum_{i-j} (r_{ij} - w_D)E_{ij} + w_D \sum_i V_i - \chi$, and the code's partition function is therefore $Z_C = \sum_{E,V} N_{EV} \exp(-H_{EV}/w_C)$, where the summation is over all valid network configurations. The factor $N_{EV}$ is the number of possible ways to color a given pattern specified by the fields $E$ and $V$ and in general is hard to calculate. However, as common in biological codes, it is assumed henceforth that there are much more meanings than available symbols, $n_m >> n_s \geq n_f$, so that $N_{EV} \approx \exp(n_f \ln n_m)$. By substituting into $Z_C$ the approximated $N_{EV}$ with the island number $n_f$ taken from Eq. 3, the partition function becomes

$$(4) \qquad Z_C = \exp(-\chi / w_C) \sum_{E,V} \left( \prod_{i-j} \exp(-\beta_{ij} E_{ij} / w_C) \prod_i \exp(-\alpha V_i / w_C) \right),$$

with the coefficients $\alpha = w_D + w_C \ln n_m$ and $\beta_{ij} = (r_{ij} - w_D) - w_C \ln n_m$.

The code partition function $Z_C$ (Eq. 4) sums over all possible networks. The building blocks of these networks are self-avoiding polymers, which fuse to each other at the junctions. These junctions cannot be regarded as the crossing points of non-interacting polymers because free "dangling ends" are forbidden and because the formation of junctions affects the total energy of the configuration. Within the analogy to physical networks, Eq. 4 may seem as a summation over two chemical "species" – one which resides on the edges with "excitation energies" (or chemical potentials) $\beta_{ij}$, and one on the vertices with the excitation energy $\alpha$. However, these two species are not really independent and cannot be summed separately. A vertex is occupied if and only if its coordination number is at least two because "dangling-ends" or isolated vertices are forbidden. Similarly, an edge is occupied if and only if it connects two occupied vertices. The relevant chemical species are the monomers, i.e pairs a vertex and a neighboring edge, that carry energies of $\alpha + \beta_{ij}$ and the possible $q$-fold junctions ($q > 2$). The formation of a $q$-junction replaces $q$ ends that contribute each an energy of $\alpha/2$ by one vertex of energy $\alpha$ so that the overall energy change is $(1 - q/2)\alpha$. Performing the summation over all possible networks proves to be tricky as the vertex and edge occupations are not independent. Below, we employ the $n = 0$ formalism to resolve this configuration counting problem.

**Correspondence between code optimization and spin networks.** The $n = 0$ formalism was devised by de Gennes to examine polymer solutions (11, 12). Recently it was applied also to microemulsions, micellar solutions and dipolar fluids (13). At the basis of this approach is a mathematical equivalence between a system of self-excluding polymers to a system of interacting $n$-component magnets, in the limit of vanishing number of components, $n = 0$ (this is a formal limit, which does not correspond to any realistic spin system). The $n = 0$ formalism is reviewed elsewhere (12, 13). Here we discuss only concisely the basic idea and use this approach to show that the statistics of the code-table (Eqs. 2, 4) can be mapped to that of the zero-component magnets.

To demonstrate the equivalence between the code-table problem and the spin system, let us consider the dual of the symbol-graph on which the polymer network resides (Fig. 1) and assign to each of the edges $i$-$j$ an $n$-component magnetic spin $\mathbf{S}_{ij}$. The interaction is represented by a spin-Hamiltonian $H_S$ and the spin partition function is $Z_S = \langle \exp(-H_S) \rangle$, where $\langle \ldots \rangle$ denotes the average over all possible spin orientations. A peculiar feature of this system in the limit of zero components ("the $n = 0$ property") is that all the averages over products of spins vanish except for the quadratic averages $\langle S_{ij}^2 \rangle = 1$, where $S_{ij}$ is any of the components of $\mathbf{S}_{ij}$. This property enables mapping of the spin lattice to the network ensemble by tailoring a spin Hamiltonian $H_s$ and a consequent spin partition-function $Z_S$, in which an $\langle S_{ij}^2 \rangle$ term appears *if and only if* the corresponding edge $i$-$j$ is occupied in the network partition function $Z_C$. It is shown below that this correspondence is accomplished by the Hamiltonian $H_s$,

$$(5) \qquad H_S = -\sum_i h_i, \quad h_i = a_i \left[ \prod_{j(i)} \left(1 + b_{ij} S_{ij}\right) - \sum_{j(i)} b_{ij} S_{ij} - 1 \right],$$

where $j(i)$ are all the neighbors of $i$ and the coefficients $a_i$ and $b_{ij}$ are related to the $\alpha$ and $\beta_{ij}$ parameters (Eq. 4). The functions $h_i$, the contributions of each vertex $i$ to $H_s$, consists of all possible products of two or more edges around $i$. Each of these products corresponds to a possible edge occupation state in a network (Fig. 2). The one- and zero-edge configurations are forbidden in the network and are therefore subtracted from $h_i$ (the second and third terms).

How this form of the Hamiltonian $H_s$ ensures the equivalence is clear by expanding $Z_S$ in a power series, $Z_S = \langle \exp(-H_S) \rangle = \langle \Pi_i \exp(h_i) \rangle = \langle \Pi_i (1 + h_i + h_i^2/2 + \ldots) \rangle$. Due to the $n = 0$ property, the infinite series can be exactly truncated at the second term and we obtain $Z_S = \langle \Pi_i (1 + h_i) \rangle$, which is a sum over averages of spin products. In this sum, the only non-vanishing terms are those in which each spin $S_{ij}$ appears *exactly* twice. In those configurations, $S_{ij}$ must appear in both contributions of $h_i$ and $h_j$. As illustrated in Fig. 2, it is evident that such a spin configuration corresponds to a network configuration and the weight of this term in the partition function is a product of the $b_{ij}$-s and the $a_i$-s of the occupied edges and vertices. From all this we find that the spin partition function is

$$(6) \qquad Z_S = \langle \exp(-H_S) \rangle = \sum_{E,V} \prod_{i-j} b_{ij}^{2E_{ij}} \prod_i a_i^{V_i},$$

with the same occupation variables $E_{ij}$ and $V_i$ which are used to count network configurations in the code partition function $Z_C$ (Eq. 4). Finally, to obtain the one-to-one correspondence one needs to identify the "fugacities" in Eqs. 4 and 6, $b_{ij}^2 = \exp(-\beta_{ij}/w_C)$ and $A_i = \exp(-\alpha/w_C)$, which results in identical partition-functions, $Z_C = Z_S$ (up to an irrelevant factor). In the following, this correspondence is used to gain insight into the noisy coding system from a mean-field solution of the spin system.

**Mean-field solution of the n = 0 spin system.** To solve for the optimum of a noisy molecular code, we employ a standard variational mean-field approach (for details see Methods). First, we approximate the spin probability distribution by a product of independent single-spin distributions, $\rho = \Pi_{ij} \rho_{ij}(S_{ij})$, which disregards the correlation among the spins. Then, using this mean-field distribution we calculate a variational least upper bound, $F_M$, on the free energy $F_S$ of the system, $F_M \geq F_S = -\ln Z_S$. By this mean-field procedure it is straightforward to find that the average spin, $s_{ij} = \langle S_{ij} \rho_{ij}(S_{ij}) \rangle$, is given by the relation

$$(7) \qquad s_{ij} = \frac{g_{ij}}{1 + \frac{1}{2} g_{ij}^2},$$

where the $g_{ij}$ are effective fields that involve only the average spins on neighboring edges, $g_{ij} = a_i b_{ij} [\Pi_{k(i) \neq j}(1 + b_{ik} s_{ik}) + \Pi_{l(j) \neq i}(1 + b_{jl} s_{jl}) - 2]$ (Methods). In a similar fashion we obtain the mean-field free energy, $F_M = -\sum_i h_i + \sum_{i-j} [g_{ij} s_{ij} - \ln(1 + g_{ij}^2/2)]$. The first term in $F_M$ is the average Hamiltonian, which is the quality of the average code, while the last term is entropic and accounts for the cost of the code. The self-consistency relations of Eq. 7 are polynomial equations in the average spins, which link every spin $s_{ij}$ to the spins on the neighboring edges. Although in the general case a solution is obtained only numerically, it is much simpler to solve than the typical rate-distortion expression (e.g. (*6*)). More important, it provides insight into the "low-temperature" (low $w_C$) regime where the landscape of the code free energy is rugged and therefore hard to calculate.

To make use of the equivalence between the spin system and the coding network, we need to express the average network occupancies, $e_{ij} = \langle E_{ij} \rangle$ and $v_i = \langle V_i \rangle$ as a function of the average spin $s_{ij}$. The average edge-occupancy $e_{ij}$ is given by $e_{ij} = (1/2)\partial \ln Z_S / \partial \ln b_{ij} = g_{ij} s_{ij}/2$, a consequence of Eqs. 5-6. Likewise, the average vertex-occupancy is $v_i = \partial \ln Z_S / \partial \ln a_i = h_i$ (Methods). Thus, one can calculate the average network configuration (that is the average code) for any value of the fugacities $A_i$ and $B_{ij}$, or for the equivalent control parameters of the coding system, $w_D$, $w_C$, $r_{ij}$ and $n_m$. This is demonstrated below, where a first-order "coding transition" is deduced from the spin formalism.

Mean-field models similar to the one used here are standard in n = 0 treatments of self-assembling systems, such as polymer and micellar solutions (14, 15) and networks (13). The basic idea of the mean-field approach is to replace complicated many-spin interactions by an effective interaction of a single spin with an effective field (the $g_{ij}$ polynomials). This procedure vastly simplifies the problem and enables a relatively simple solution. However, this simplicity comes at the cost of disregarding the long-range spatial correlations between the spins and the corresponding correlations between the edges and the vertices. For example, one can estimate the mean connectivity in the network, but cannot tell how many loops it contains. In general, a mean-field treatment merely approximates qualitatively the behavior of thermodynamic functions. However, the accuracy of this approximation improves when each spin interacts with many neighbors. Therefore, when the symbol graph is highly connected – such as the graph of the genetic code, where each codon has nine neighbors – the mean-field approximation is expected to be relatively accurate and provides a basis to more elaborate models.

**A first-order coding transition.** The formal equivalence of the spin system and the code-table enables us to follow the evolution of the code-table in response to variation of the control parameters which govern its optimization: the cost weight $w_C$, the misreading matrix $r_{ij}$, the diversity weight $w_D$ and the number of meanings $n_m$, which measures the richness of the meaning space. These parameters are not independent and are related through the spin fugacities, $a$ and $b_{ij}$. It proves convenient to represent these relations in terms of the normalized diversity, $w_D^* \equiv w_D/w_C + \ln n_m = -\ln a$, and the normalized misreading probabilities, $r_{ij}^* \equiv r_{ij}/w_C = -\ln(ab_{ij}^2)$.

To examine the response of the code-table to variation of the four control parameters, we consider, for the sake of simplicity, regular (or isotropic) symbol graphs, in which all the vertices and edges are equivalent. Regular symbol graphs are useful in biological context. For example, a regular graph may approximate the symbol graph of the genetic code, where each of the sixty-four codons has nine neighbors (2). Regular graphs may also describe large symbol spaces, for example the space of DNA binding site, whose structure is not exactly known but only the average coordination number. Regularity of the symbol graph implies uniform misreading $r_{ij} = r$ and, as a result, homogenous average spin $s_{ij} = s$, edge occupancy $e_{ij} = e$ and spin occupancy $v_i = v$. Due to symmetry, we need to solve a single self-consistence relation $s = g/(1 + g^2/2)$ (Eq. 7) with $g = 2ab((1 + bs)^{q-1} - 1)$. Similarly, the free energy per vertex of the dual graph is given by $F_I = -h + (q/2)gs + (q/2)\ln(1 + g^2/2)$, where $h = a((1 + bs)^q - qbs - 1)$ and $q$ is the coordination number.

The resulting phase diagram of the regular symbol graph exhibits a line of first-order transitions, where the number of encoded meanings jumps discontinuously from $n_f = 1$, that is a mapping encodes a sole meaning, to a number $n_f > 1$ that scales extensively with the size of the symbol-graph $n_s$ (Fig. 3). The state $n_f = 1$ is termed non-coding, since the coding system at this state conveys no information as only one symbol is used. When a coding state, $n_f > 1$, emerges the coding system is capable of conveying information at the rate of $\log_2 n_f$ bits/symbol. Tracing the behavior of the free energy $F_I$ as the scaled misreading $r^*$ is varied at a constant scaled diversity $w_D^*$, we find that at high $r^* = r_{ij}/w_C$ there is no network. This is because the system prefers to reduce the impact of misreading errors, which are too costly at a high $r^*$, at the expense of diversity. In the free energy curve, this is manifested by a global minimum at $s = e = v = 0$.

As $r^*$ decreases the system reaches the first-order transition. The transition line describes an increasing concave curve in the $w_D^*-r^*$ plane (Fig. 3B). This line indicates various pathways that the system can take towards the formation of a network. The transition line can be approached when the diversity becomes more important with respect to the error-load (high $w_D^* = w_D/w_C + \ln n_m$) by (i) increasing the number of available meanings $n_m$ (ii) increasing the diversity parameter $w_D$ (iii) decreasing the misreading rate $r$ and (iv) if the system is close enough to the coding transition, $w_C$.

At the transition, the number of meaning islands jumps abruptly and becomes proportional to the number of symbols (Fig. 3C). The number of islands, i.e. the number of encoded meanings is given by Euler's characteristic (Eq. 3) with the average edge occupancy $e = gs/2$ and vertex occupancy $v = h$. For a regular graph, we find that the number of meanings per symbol is $n_f/n_s = \chi/n_s + (p/2)e - (p/q)v$, where $p$ is the coordination number in the symbol graph and $q$ in the dual graph. In this expression, $p/2$ is the number of edges per vertex (i.e. symbol) and $p/q$ is the number of cells per vertex. Next, we examine possible effects of population dynamics on the evolution of the code.

**Discussion**

The mean-field solution allows us to draw an approximate fitness landscape where the evolution of the code takes place. In general, this code fitness landscape $F_C(s_{ij})$ resides in highly-dimensional *code space*, whose coordinates are the average spins $s_{ij}$ or their conjugates, the average edge occupancies $e_{ij}$. Each point in this space is a vector **s** = $(s_{ij}, s_{kl}...)$ of dimension equal to the total number of edges, which represents a possible code. Symmetry may reduce this dimensionality and for the regular symbol-graph it becomes one-dimensional (Fig. 3). We imagine a population of "organisms", simple information-processing systems, which compete according to the fitness of their codes. Each organism is depicted as a point in code space positioned at its code **s** and the population is described by a probability density $\Psi(\mathbf{s})$. The preceding discussion assumed implicitly that, as the control-parameters change, the evolution of the code more or less follows the track of the optimum in code space. In other words, the population density is a delta-distribution located at the optimal $F_C$. We conclude this work, by considering several more realistic scenarios. First, we discuss the possibility that the coding system is stuck at metastable sub-optimal states. Then, we consider mutations and genetic drift that may broaden the population towards codes of lesser fitness.

**Metastable sub-optimal codes.** Even when the global optimum in the code fitness is at a network state of non-zero spin $s_C \neq 0$, the no-network $s = 0$ state may remain locally stable for some parameter range (Fig. 3B). This local stability means that the code is optimal with respect to small variations **δs** in the code, which may result, for example, from small changes in binding energies due to point mutations in binding sites. Metastable states may be the relevant ones on evolutionary time-scales that are short compared to the typical time it takes the system to climb the free-energy barrier towards the global minimum. For example, the evolutionary dynamics is expected to be relatively slow when the mutation rate is low, or when the population is large and the diffusion in the fitness landscape via genetic drift (random reproduction fluctuations) is therefore slow.

To locate the metastable state, one examines the curvature of the free energy at its $s = 0$ extremum, which in the case of an isotropic symbol graph is $F_I'' = \partial^2 F_I / \partial s^2 \sim 1 - 2(q-1)\exp(r^*)$. A metastable state exists as long as this curvature is positive. We find that the curvature changes its sign at the line $r^* = r/w_C = \ln(2q - 2)$, which corresponds to a "monomer" (a vertex plus an edge) of energy $\alpha + \beta = w_C \ln(2q - 2)$. This may be further clarified by considering the limit of the vertex/edge occupancy ratio near the no-network state, $v/e \rightarrow q/2$. Since there are $q/2$ edges per vertex, this indicates that at the $s = 0$ limit the dominant building block of the dilute network is the monomer. The coding system becomes unstable exactly when the monomer energy decreases below a critical value proportional to $w_C$, the "thermal energy".

**Effects of mutations and genetic drift in the codes space.** So far, it was assumed that the population of organisms is sharply concentrated around a certain optimal or metastable, sub-optimal code. This scenario applies to large populations at negligible mutation rates $\mu$. Let us consider two possible effects of population dynamics that may smear the population over the code space, mutations and genetic drift. These effects were analyzed in detail within the framework of rate-distortion theory elsewhere (6) and are discussed here only schematically, in the context of the present polymer network model.

We consider a population that is localized around a fitness optimum $F^*$ at an optimal code $s^*$, where the landscape is approximately $F_C(s) \approx F^* + \frac{1}{2}F''(s - s^*)^2$. Mutations drive the organisms to diffuse into regions of code space where the fitness is somewhat lower. This effect may be described in terms of reaction-diffusion dynamics of the probability distribution, $\partial_t \Psi(s) = \mu \cdot \partial_s^2 \Psi(s) + F_C(s)\Psi(s)$, where the first term is due to mutations and the second represents reproduction at a rate $F_C(s)$. It is straightforward to find that this dynamics tends to a steady-state, in which the population is broadened into a Gaussian, $\Psi(s) \sim \exp[-(F''/2\mu)^{1/2}(s - s^*)^2]$ (6). The width of this Gaussian scales like $\sim (\mu/F'')^{1/4}$, which indicates that mutation-induced broadening may be significant even at moderate mutation rates $\mu$. It also implies that the width of a population at a metastable state ($s = 0$) will diverge near the metastability limit ($F'' = 0$) just before the Gaussian migrates to the coding state ($s_C \neq 0$).

When the effective population size $N$ is relatively small, fluctuations in the reproduction rate, termed genetic drift, become significant. In this regime, the dynamics is characterized by long periods when the population is localized around a fitness optimum, which are separated by fast diffusive migrations to new optima. The dynamics of the distribution is known to reach a steady-state Boltzmann partition, $\Psi(s) \sim \exp(-NF_C(s))$, where the population size plays the role of an inverse temperature (6, 16). Relatively small populations (which are "hot" in this sense) are expected to be partitioned by genetic drift between the available fitness optima. For example, the two minima of the isotropic free energy $F_I$ (Fig. 3) will be populated according to their fitness values.

In the previous sections we have seen that each organism experiences "internal" noise due to stochastic molecular recognition in its coding system. The internal noise affects the fitness of the code through the error-load and the cost. On top of this, mutation and genetic drift add "external" sources of noise, which may drive parts of the population away from the optimal code. In additon, the existence of metastable states may delay the transition to a coding state. The present model and its conclusions suggest that the $n = 0$ polymer network formalism is a potential tool to study other aspects of noisy coding systems.

**Methods**

**The cost of a code-table.** The cost of a code-table is traditionally measured by the mutual information $I$ between the symbols and the meanings that they encode, $I = S(\text{meanings}) + S(\text{symbols}) - S(\text{meanings, symbols})$, where $S(\text{meanings})$ and $S(\text{symbols})$ are the entropies of the meaning space and symbol space, respectively, and $S(\text{meanings, symbols})$ is the joint entropy of these two spaces. The entropy of meanings $S(\text{meanings})$ is determined by their given distribution $P_\alpha$, $S(\text{meanings}) = -\sum_\alpha P_\alpha \ln P_\alpha$, and similarly, the entropy of symbols is $S(\text{symbols}) = -\sum_i P_i \ln P_i = \ln(n_s)$. We may therefore neglect these constant terms in the cost $I$ and consider only the joint entropy $S(\text{meanings, symbols})$, which is the term that can be optimized by tuning the average partition pattern $e_{ij}$. The joint entropy is simply the entropy of all the possible coloring patterns as determined by all possible networks and the number of possible ways to color every such pattern. It follows that the cost is therefore minus the coloring entropy $I = -S(\text{meanings, symbols}) = -S_C$.

**Graph embedding and the dual graph.** The embedded graph divides the surface into faces or cells, hexagons in our example (Fig.1). Then, one finds the dual by the following correspondence (10): Every vertex in the symbol graph corresponds to a cell in the dual (a triangle in this example) whereas every cell in the symbol graph (a hexagon) corresponds to a vertex of the dual. The correspondence between the edges in the symbol graph and its dual is one-to-one; every edge corresponds to the edge that crosses it in the dual. The resulting dual graph is a triangular lattice. The hexagonal lattice is a regular graph in which all the vertices have the same coordination number. However, the embedding procedure described here applies to any connected graphs whether it is regular or not.

**The mean-field approximation.** Within this approximation, the spin probability distribution decouples into a product of independent single-spin distributions, $\rho = \Pi_{ij}\,\rho_{ij}(S_{ij})$. To perform the mean-field calculation, we use a variational inequality, which sets an upper limit $F_M$ on the spin free energy, $F_S \leq F_M = \langle \rho H_S \rangle + T\langle \rho \ln \rho \rangle$, where $\rho$ satisfies probability conservation, $\langle \rho \rangle = 1$. We augment $F_M$ with a Lagrange multiplier to account for probability conservation, $L_M = F_M + \eta\langle\rho\rangle$, and take the functional derivative $\delta L_M/\delta\rho_{ij} = 0$. The resulting distributions are $\rho_{ij}(S_{ij}) = \exp(g_{ij}S_{ij})/\langle\exp(g_{ij}S_{ij})\rangle$, where the effective fields $g_{ij}$ involve only the average spins on neighboring edges, $g_{ij} = \partial H_S/\partial S_{ij} = \partial(h_i + h_j)/\partial S_{ij} = a_i b_{ij}[\Pi_{k(i)\neq j}(1 + b_{ik}s_{ik}) + \Pi_{k(j)\neq i}(1 + b_{jk}s_{jk}) - 2]$ with $s_{ij} = \langle\rho S_{ij}\rangle = \langle S_{ij}\rho_{ij}(S_{ij})\rangle$. From the $n = 0$ property it follows that $\langle\exp(g_{ij}S_{ij})\rangle = \sum_k g_{ij}^k \langle S^k\rangle/k! = 1 + g_{ij}^2/2$, which leads to $\rho_{ij}(S_{ij}) = \exp(g_{ij}S_{ij})/(1 + g_{ij}^2/2)$. As usual in the mean-field approach, the solution for the average spins is derived from the self-consistency conditions, $s_{ij} = \langle S_{ij}\rho_{ij}(S_{ij})\rangle$, which together with the spin distribution yield the relations, $s_{ij} =$ $\langle S_{ij}\exp(g_{ij}S_{ij})\rangle/(1 + g_{ij}^2/2) = g_{ij}/(1 + g_{ij}^2/2)$ (Eq. 7). For each edge there corresponds one relation, so that overall there are $n_e$ relations. In a similar fashion we obtain the mean-field free energy, $F_M = -\sum_i h_i + \sum_{(i,j)} g_{ij}s_{ij} - \sum_{(i,j)} \ln(1 + g_{ij}^2/2)$, and it is easy to verify that the relations define the extremum $\partial F_M/\partial s_{ij} = 0$. Eq. 7 is analogous to the familiar self-consistency relation of the mean-field Ising magnet, $s = \sinh(gs)/\cosh(gs)$; the different form is due to the truncation of the power-series expansion of the hyperbolic sine and cosine thanks to the $n = 0$ property. Solving (7) for $g_{ij}$ as a function of $s_{ij}$, we find that solution can be described graphically as the points where the function $e_{ij} = g_{ij}s_{ij}/2$ crosses the ellipse $2s_{ij}^2 + (2e_{ij} – 1)^2 = 1$ (Fig. 3A).

**The use of Euler's characteristic.** When we apply the Euler characteristic (Eq. 3), an underlying assumption is that the embedding is cellular (10), i.e. every meaning island is homeomorphic to an open disc. This is not necessarily true when the number of islands is less than the number of holes in the surface, which is equal to the genus $\gamma = 1–\chi/2$. In this case, some islands are expected wrap between two holes and therefore are not homeomorphic to a disk. However in the "thermodynamic limit" of many islands per hole, $n_f \ll |\chi|$, the embedding is mostly cellular and is well approximated by Eq. 3.

**Figure legends**

Fig. 1. **The code-table as an information channel and the relation to polymer networks.** The code-table is an information channel or mapping that relates a space of meanings (left), which are depicted as colors), to a symbol space. The symbol space (middle) is a graph, hexagonal in this example, in which vertices are symbols and edges connect symbols that are likely to be confused by reading. The code-table induces a coloring pattern of the meaning space, which divides it into meaning islands (dotted lines). In the dual graph, triangular in this example (right), the boundaries between the meaning islands form a network of "polymers" (solid lines).

Fig. 2. **Correspondence of the polymer networks to $n = 0$ spin systems.** The solid lines denote boundaries between the meaning islands induced by the code on the dual graph (Fig. 1, right). In the spin model, each edge is assigned with a spin $S_{ij}$. Each vertex $i$ contributes to the spin Hamiltonian $H_S$ a factor $h_i$, which accounts for all possible edge occupancies around this vertex. By the construction $h_i$ (Eq. 5), if a vertex is occupied then at least two of the adjacent edges are occupied. In the present example, a four-junction at vertex 1 (red), which corresponds to a factor $a_1 b_{13} S_{13} b_{14} S_{14} b_{16} S_{16} b_{17} S_{17}$, connects to three linear elements (magenta), e.g. $a_7 b_{71} S_{71} b_{79} S_{79}$, and one three-junction (green), $a_3 b_{31} S_{31} b_{32} S_{32} b_{38} S_{38}$. The corresponding contribution to the spin partition function $Z_S$ is an average over all the spin orientations. This contribution does not vanish because each spin appears exactly twice in the product since $S_{ij}$ appears exactly once in both edge configurations of $i$ and $j$. The weight of this contribution is a product of $b_{ij}$-s and $a_i$-s for each edge and vertex in the product.

Fig. 3 **The free energy and phase diagram of the code-table.** (A) The free energy $F_I$ (top) and the edge occupancy $e$ (bottom) of the isotropic hexagonal symbol graph (Fig. 1) at scaled diversity $w_D^* = 1.77$ and scaled misreading $r^* = 0.49$ (red), 0.47 (green), 0.36 (blue) and 0.27 (black). All the curves exhibit a minimum at the no-network state $s = 0$, $F_I(0) = 0$. At the coloring transition (green curve) the second minimum that corresponds to the network state $s_C$ is at $F_I(s_C) = F_{iso}(0) = 0$. At lower values of $r^*$ the network state becomes the global minimum. At $r^* = (2q - 2)$ the no-network state becomes unstable (black curve). The dashed curve traces the loci of the network state as $r^*$ varies. These loci are found at the intersection of the edge density $e = gs/2$ and the ellipse $2s^2 + (2e - 1)^2 = 1$ (bottom). (B) The phase diagram of the coloring (network ↔ no network) transition (solid line) in the $w_D^*$–$r^*$ plane. The dashed line bounds the region of metastability, beyond which the no-network state is destabilized. (C) The vertex occupancy $v$ (blue), edge occupancy $e$ (black) along the transition line (B). The ratio meanings-per-symbol $n_I/n_s$ is calculated from the occupancies using Euler's characteristic (Eq. 3).

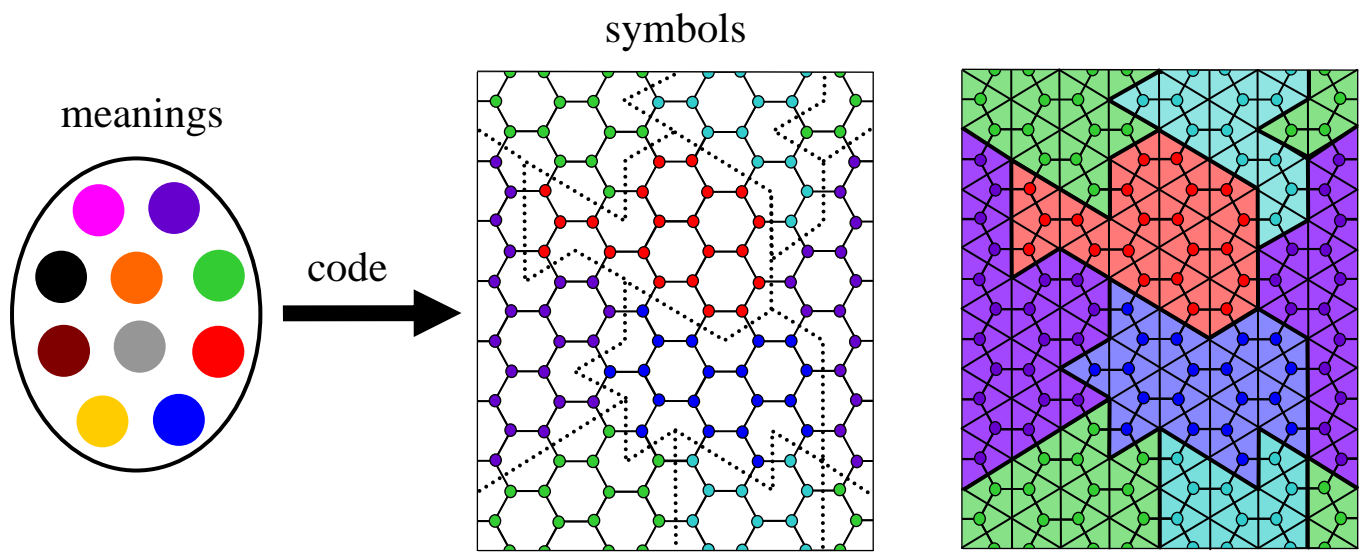

Figure 1

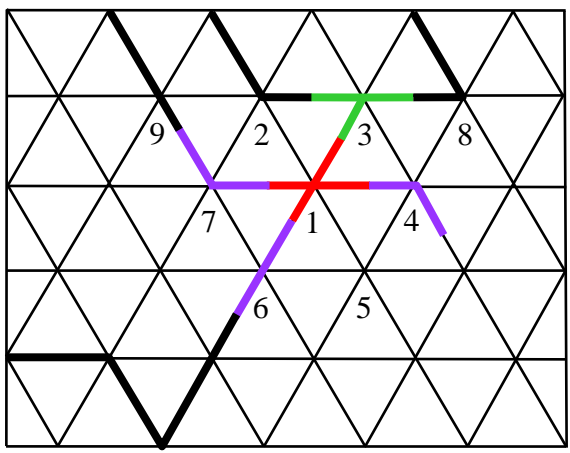

Figure 2

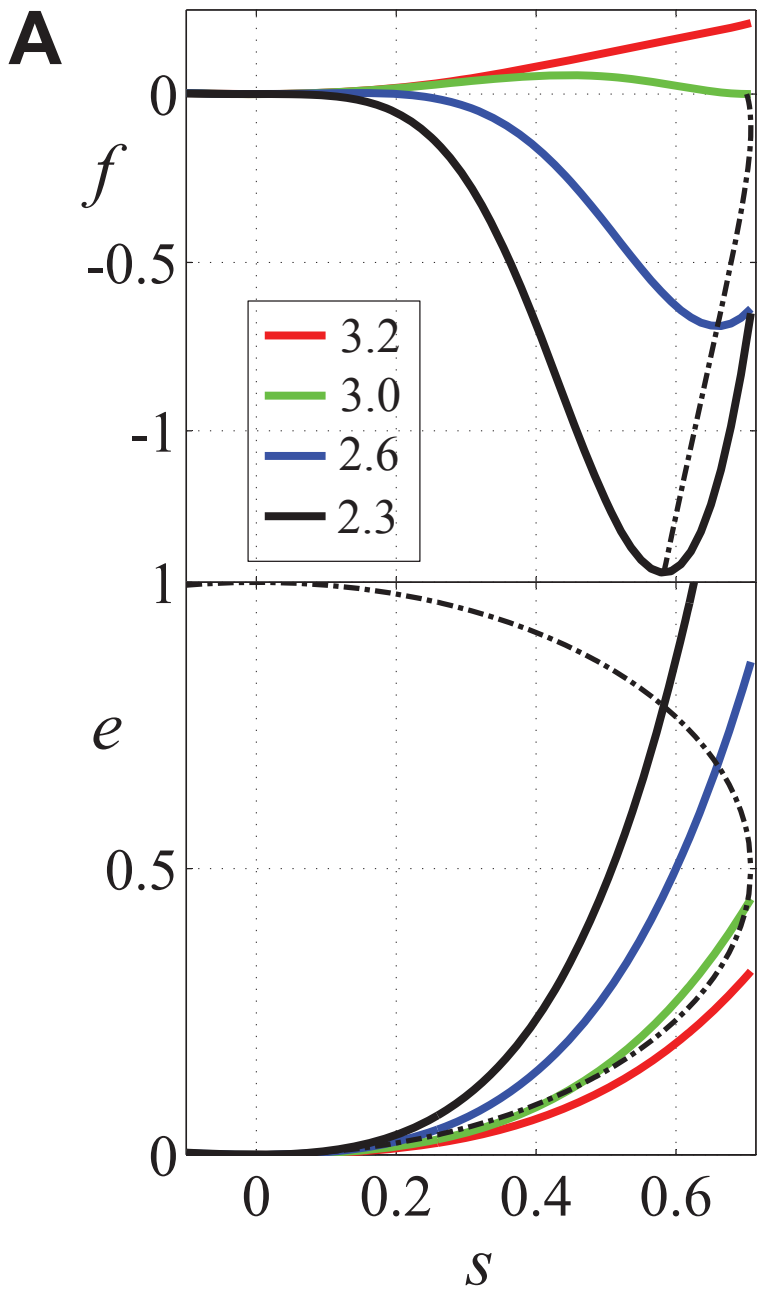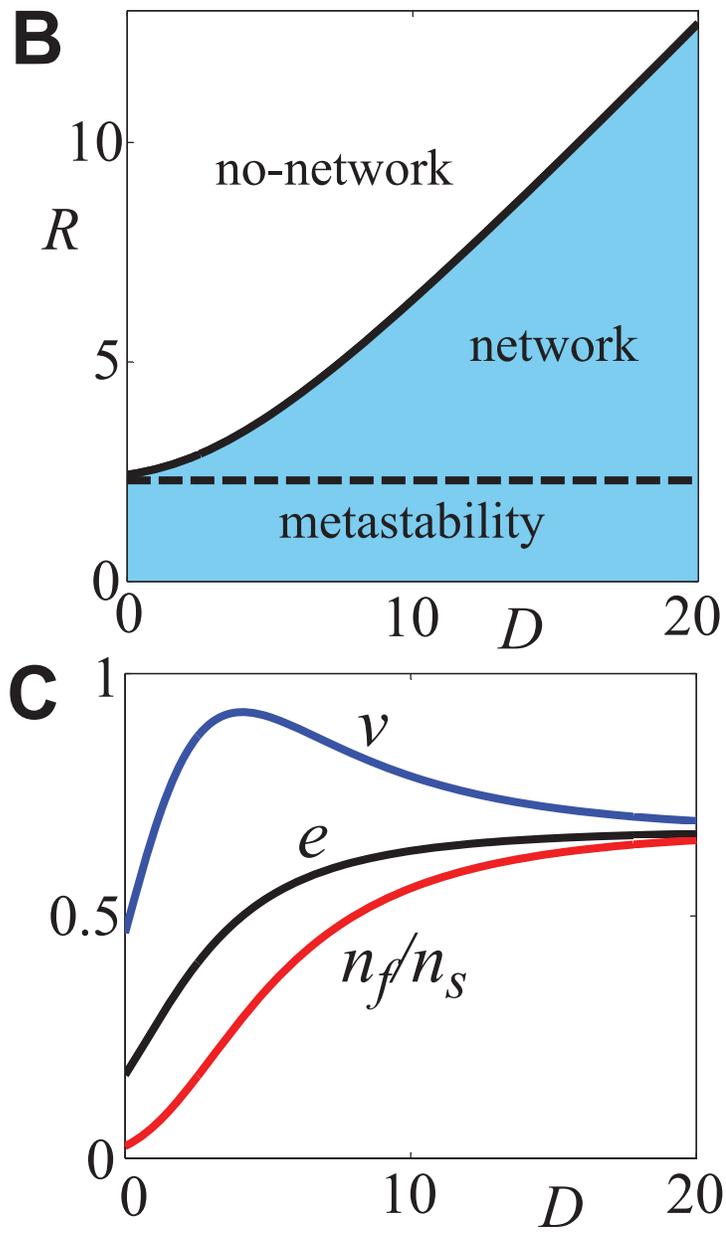

Figure 3